# TAROT: Towards Optimization-Driven Adaptive FEC Parameter Tuning for Video Streaming

Jashanjot Singh Sidhu[*]
Concordia University, Canada
jashanjot.sidhu@mail.concordia.ca

Aman Sahu[*]
Concordia University, Canada
aman.sahu@mail.concordia.ca

Abdelhak Bentaleb
Concordia University, Canada
abdelhak.bentaleb@concordia.ca

**Notice:** This manuscript is an **extended version** of our accepted paper at the **17th ACM Multimedia Systems Conference (ACM MMSys 2026)**.

## ABSTRACT

Forward Error Correction (FEC) remains essential for protecting video streaming against packet loss, yet most real deployments still rely on static, coarse-grained configurations that cannot react to rapid shifts in loss rate, goodput, or client buffer levels. These rigid settings often create inefficiencies: unnecessary redundancy that suppresses throughput during stable periods, and insufficient protection during bursty losses, especially when shallow buffers and oversized blocks increase stall risk. To address these challenges, we present TAROT, a cross-layer, optimization-driven FEC controller that selects redundancy, block size, and symbolization on a per-segment basis. TAROT is codec-agnostic—supporting Reed–Solomon, RaptorQ, and XOR-based codes—and evaluates a pre-computed candidate set using a fine-grained scoring model. The scoring function jointly incorporates transport-layer loss and goodput, application-layer buffer dynamics, and block-level timing constraints to penalize insufficient coverage, excessive overhead, and slow block completion. To enable realistic testing, we extend the SABRE simulator [1] with two new modules: a high-fidelity packet-loss generator that replays diverse multi-trace loss patterns, and a modular FEC benchmarking layer supporting arbitrary code/parameter combinations. Across Low-Latency Live (LLL) and Video-on-Demand (VoD) streaming modes, diverse network traces, and multiple ABR algorithms, TAROT reduces FEC overhead by up to 43% while improving perceptual quality by 10 VMAF units with minimal rebuffering, achieving a stronger overhead–quality balance than static FECs.

---

[*]The first two authors contributed equally to this work.
[1]TAROT code is available at: https://github.com/IN2GM-Lab/TAROT-FEC

## 1 INTRODUCTION

The rapid growth of video traffic [46] places unprecedented pressure on content providers to sustain consistently high Quality of Experience (QoE), especially as 4K and high–frame-rate formats become mainstream. Delivering such content over today's heterogeneous access networks—particularly LTE and 5G—remains challenging. Despite offering high peak throughput, these networks exhibit sharp short-term fluctuations, sporadic congestion, and non-trivial packet loss [3, 13, 28, 54], all of which deteriorate the effective goodput available to video players. As a result, maintaining stable, high-quality playback in the presence of such volatility has become increasingly difficult.

Historically, most QoE optimization efforts have focused on two layers of the streaming stack. At the *client side*, HTTP Adaptive Streaming [7] (HAS) algorithms select the "best" bitrate representation based on observed bandwidth and buffer conditions. At the *transport side*, congestion-control [29] mechanisms shape traffic to avoid queue buildup and adapt sending rate to network capacity. While these mechanisms are effective under bandwidth variability, they remain fundamentally ill-suited for lossy environments. Even modest packet loss can trigger superlinear goodput collapse, degrade throughput estimates, and cause ABR algorithms to oscillate or fall back to low-quality representations, regardless of the available peak bandwidth. These challenges are amplified in volatile LTE, and 5G environments, where retransmissions become counterproductive—increasing latency, reducing goodput, and risking a congestion-loss spiral.

This mismatch has motivated the adoption of forward error correction [18, 44] (FEC) as a complementary reliability mechanism for video delivery. By proactively injecting repair symbols, FEC allows receivers to recover lost data without depending on transport-level retransmissions. In principle, this should reduce playback stalls, improve resilience to burst losses, and stabilize representation selection within HAS clients. However, practical deployments of FEC in streaming systems suffer from several longstanding limitations. Most existing systems [23, 49, 59] rely on *static* FEC parameters—fixed block sizes, fixed symbol sizes, and fixed redundancy ratios—that remain unchanged throughout the streaming session. Such configurations are easy to deploy but brittle in practice. When the network is stable, static FEC wastes bandwidth and reduces the goodput available to the ABR logic. When the network becomes lossy, the same fixed redundancy often proves insufficient, causing decoding failures and triggering retransmissions that defeat the purpose of FEC.

To overcome these limitations, several heuristic and learning-based redundancy controllers have been proposed. Heuristic approaches [14] adjust the redundancy ratio based on observed packet loss or estimated congestion, but they typically treat FEC as a single scalar "overhead knob" and cannot adapt the underlying block structure. Recent reinforcement-learning approaches [6, 26] offer more flexibility but remain constrained by a fixed coding design and do not account for segment-level timing constraints, buffer state, or the non-linear collapse of effective throughput under loss. As a result, existing solutions struggle to balance recovery capability and overhead efficiency across diverse network conditions, particularly in today's highly variable 5G environments.

These challenges expose a fundamental gap in the design of reliability mechanisms for video streaming: FEC must adapt not only *how much* redundancy is added, but also *how* it is structured, *when* it should be applied, and *how* it interacts with buffer dynamics, segment duration, and loss-induced goodput degradation. Addressing this gap requires a fine-grained, cross-layer approach capable of making per-segment decisions based on real-time transport and playback signals. To address these limitations, we propose TAROT[2], a fine-grained, cross-layer FEC controller designed specifically for HAS video streaming over volatile networks. TAROT departs from conventional redundancy controllers by operating at the *segment level* and by jointly reasoning about packet loss, redundancy overhead, encoding latency, and segment-timing constraints. Instead of treating FEC as a single scalar overhead knob, TAROT evaluates multiple code families and block configurations $(n, k, S)$—where $n$ denotes the number of source symbols, $k$ the number of repair symbols, and $S$ the symbol size—for every segment and selects the configuration that minimizes a multi-objective cost function balancing overhead, recovery capability, and perceptual impact.

To support realistic evaluation of such per-segment decisions, we extend the widely adopted SABRE [53] simulator and develop *SABRE–FEC*, a modular packet-level FEC framework that models both redundancy-induced bandwidth inflation and the computational cost of generating repair symbols. SABRE–FEC integrates an analytical loss model that captures the non-linear collapse of goodput under packet loss, and incorporates encoding latency directly into the segment download time. This enables reproducible, controlled, and implementation-agnostic evaluation of fine-grained FEC strategies within HAS logic. The key contributions of this paper are three-fold:

(1) We develop TAROT, a per-segment adaptive FEC controller that formulates FEC parameter selection as a constrained optimization problem and solves it efficiently via candidate enumeration. TAROT jointly selects the FEC code family and configuration $(n, k, S)$ for every segment, accounting for overhead, encoding latency, recovery capability, and segment deadlines through a lightweight multi-objective scoring model, enabling robust operation under highly variable LTE and 5G conditions.

(2) We build SABRE-FEC, an extension of the SABRE simulator with a modular packet-level FEC layer and an analytical loss model. SABRE-FEC realistically models transmitted-byte inflation and encoding latency, enabling controlled and reproducible evaluation of fine-grained FEC strategies.

(3) We conduct a comprehensive evaluation of TAROT-enhanced FEC using diverse real-world traces (LTE, 5G) and both LLL and VoD streaming modes. Our results show that TAROT consistently improves perceptual quality and reduces overhead compared to static FEC baselines.

The rest of the paper is organized as follows. Section 2 presents related work, followed by design of TAROT in Section 3. Section 4 highlights performance evaluation of TAROT-enhanced FEC. Section 5 discusses the limitations, and Section 6 concludes the paper.

## 2 RELATED WORK

**Heuristic-based FEC.** Traditional FEC mechanisms such as Reed-Solomon [23], RaptorQ [1, 59], and XOR-based erasure codes [49] have been widely deployed in multimedia systems. RTP/RTCP-based FEC [47], including the standardized parity schemes in RFC 5109 [27] also follow rule-based redundancy assignment.

**Learning-based FEC.** More recent work introduced redundancy adaptation through network telemetry or reinforcement learning. RL-FEC [61] adjusted redundancy based on measured loss or delay, while RL-AFEC [9] learned a frame-level redundancy policy using RL and VMAF-based feedback. DeepRS [11] and LightFEC [21] used LSTM-based packet-loss prediction to tune RS redundancy dynamically, whereas RL-FEC for NTNs [61] learned when to inject repair packets under delayed feedback. P-FEC [60] combined intra- and inter-generation coding to meet SLA decoding guarantees under uncertain loss prediction. R-FEC [26] jointly optimized video and FEC bitrates for WebRTC using RL, and ML-powered WebRTC FEC [17] used supervised learning to improve energy-efficient adaptation on mobile devices. Tooth [6] targeted cloud gaming with a per-frame predictor that assigned redundancy based on frame length and loss aggregation. Although these methods incorporated adaptivity, they all operated within fixed coding structures and optimized only a single redundancy knob. None jointly considered the FEC family, block parameters $(n, k, S)$, encoding latency, and segment timing constraints, and all were developed for RTC or cloud-gaming workloads rather than HAS.

**SABRE simulator for HAS.** SABRE [53] is widely used to study bitrate adaptation in HTTP Adaptive Streaming. It models throughput, RTT, and buffer evolution but does not natively support packet-level loss, redundancy overhead, or FEC processing latency.

Prior FEC systems either relied on static parameters or adapted only the redundancy ratio, while existing simulators lack packet-level FEC modeling and fail to capture the interaction between overhead, encoding latency, and HAS buffer dynamics. TAROT addresses these limitations by jointly optimizing the FEC family and block configuration $(n, k, S)$ on a per-segment basis, incorporating redundancy overhead, encoding cost, loss-induced goodput collapse, and segment timing constraints into a unified decision process. This enables TAROT to provide fine-grained, adaptive FEC tailored to the rapid variability of modern LTE and 5G networks—capabilities that were absent from all prior heuristic and learning-based approaches.

---
[2]TAROT is an abbreviation of the following full name of the design: **T**owards optimization-Driven **A**daptive FEC Pa**R**ameter Tuning for Vide**O** **ST**reaming

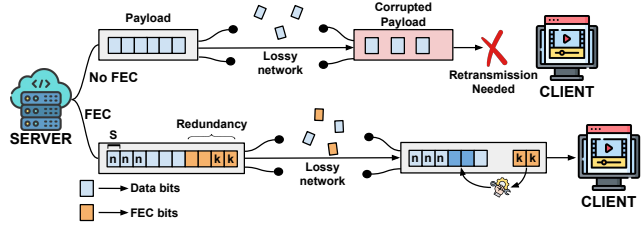

Figure 1: FEC workflow.

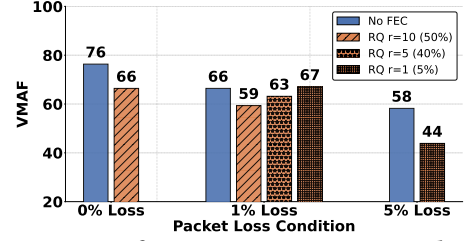

Figure 2: Impact of static FEC parameters on client QoE.

## 3 TAROT SOLUTION
### 3.1 FEC Overview

FEC provides a proactive mechanism for loss recovery by transmitting a small amount of redundant information alongside the original data. This redundancy is encoded as repair symbols, which enable the receiver to reconstruct missing portions of a packet or segment without requiring retransmissions. By trading modest overhead for timely recovery, FEC mitigates the impact of packet loss and reduces the latency penalties typically associated with retransmission-based recovery in lossy network environments. This mechanism is illustrated in Figure 1. As shown, when the server transmits data over a lossy channel, even a small amount of symbol loss can render the entire payload unusable at the receiver, thereby triggering a retransmission. Because the corrupted payload cannot be decoded or partially recovered, the client must wait for the full packet to be resent, introducing additional delay. In contrast, when FEC is applied, the repair symbols transmitted alongside the original data allow the receiver to recover the missing portions of the packet. As long as a sufficient combination of source and repair symbols is received, the client can fully reconstruct and decode the payload without requiring retransmissions.

Regardless of the coding technique employed, the fundamental operating principle of FEC remains unchanged. Let the original payload size be $P$. A FEC encoder selects a symbol size $S$ and divides the payload into $n$ source symbols, each of size $S$, so that the source block occupies $nS$ bytes. The encoder then applies a redundancy factor $r$ to determine the number of repair symbols, $k = \lceil rn \rceil$, that must be generated. While different FEC schemes may use different algorithms to produce these repair symbols, they all share the same structure: a block of $n$ source symbols is augmented with $k$ repair symbols to enable loss recovery without retransmissions.

The resulting FEC overhead follows directly from the number of repair symbols generated. Since each repair symbol has the same size $S$ as a source symbol, the total amount of redundant data added by the encoder is $kS$ bytes. Given that the original payload occupies $P = nS$ bytes, the overhead fraction introduced by FEC is therefore

$$\text{Overhead} = \frac{kS}{nS} = \frac{k}{n} = \frac{\lceil rn \rceil}{n}.$$

This expression captures the fundamental trade-off of FEC: increasing redundancy provides greater protection against loss but incurs proportionally higher overhead.

In practice, this overhead term $\frac{k}{n}$ interacts directly with link goodput, buffer state, and playback rate. As a result, selecting an appropriate redundancy level $r$ becomes a multi-dimensional optimization problem that must balance recovery capability, latency constraints, and bandwidth efficiency. Beyond overhead, the key property that enables FEC to eliminate retransmissions is its ability to recover from symbol loss. Once transmitted over the network, the receiver obtains some number of surviving source symbols $n_{\text{recv}}$ and surviving repair symbols $k_{\text{recv}}$. Successful decoding requires only that the total number of received symbols meets the feasibility condition: $n_{\text{recv}} + k_{\text{recv}} \geq n$, in which case all $n$ original source symbols can be reconstructed, regardless of which individual symbols were lost. Thus, FEC effectively masks loss as long as the redundancy is sufficient to cover the number of missing symbols.

In adaptive streaming scenarios, the redundancy requirement is tightly coupled to network and playback dynamics. Higher packet loss necessitates larger repair budgets ($k$), whereas limited link goodput or a low playback buffer constrains the amount of FEC overhead that can be tolerated. To illustrate this trade-off, we conduct a HAS video-on-demand (VoD) streaming experiment on the *NVIDIA Asteroid* [37] (NVA) 4K video using our enhanced SABRE-FEC implementation (described in Section 4.1) and the Netflix 5G [41] network trace. Further experimental details are presented in Section 4. We compare *No FEC* against the default RaptorQ (RQ) configuration, which uses fixed parameters of $n$=20, $k$=10, and $S$=64 bytes.

Figure 2 presents the resulting Video Multi-method Assessment Fusion (VMAF) [43] scores across three loss scenarios. Under 0% loss, FEC provides no benefit: the 50% overhead of the default RQ setting reduces effective bandwidth, causing VMAF to drop by 10 units relative to No-FEC–approximately 2 Just Noticeable Differences[3] (JNDs) [10]. Here, the degradation arises entirely from overhead rather than recovery.

Under 1% loss, both No-FEC and RQ degrade due to symbol loss, but the gap narrows: RQ now trails No-FEC by 7 VMAF units (1 JND). Although FEC begins to help, the default redundancy ($k$=10) still imposes excessive overhead. Reducing the repair symbols to $k$=5 (25% overhead) increases VMAF, bringing the difference with No-FEC below the JND threshold. Reducing further to $k$=1 (5% overhead) makes RQ slightly outperform No-FEC by 1 VMAF unit. Notably, overhead can be reduced either by increasing $n$ or decreasing $k$, with similar qualitative impact.

However, this does not imply that RQ with $k$=1 is universally optimal. Under 5% loss, the same low-redundancy configuration fails: it lags behind No-FEC by 14 VMAF units (more than 2 JNDs). In this high-loss regime, the redundancy is too small to cover symbol losses, and decoding frequently fails, triggering retransmissions and rebuffering. Consequently, insufficient redundancy performs worse than not using FEC at all.

---
[3]A difference of 6 VMAF units is widely considered the JND threshold, above which a quality change becomes perceptible to a typical viewer.

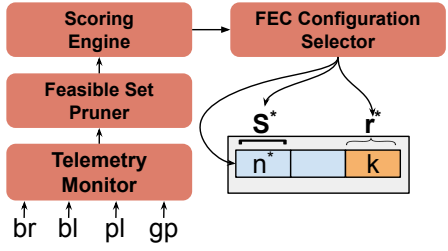

**Figure 3: TAROT adaptive FEC adjustment.**

We also evaluated the impact of increasing the symbol size $S$ to 128 and 256 bytes. Doubling $S$ approximately doubled the encoding latency. However, in a VoD setting with ample playback buffer, this computational overhead did not manifest as any perceptible quality degradation. These findings reinforce a key insight: FEC performance is highly sensitive to the interplay between redundancy, symbolization, and network conditions. No single static configuration performs well across all scenarios. Consequently, FEC must adapt its parameters dynamically—rather than relying on fixed, one-size-fits-all settings—to maintain both efficiency and robustness under varying network conditions. This key insight forms the foundation for our proposed solution, TAROT.

## 3.2 TAROT: Adaptive FEC Adjustment

Experimental analysis in Figure 2 revealed that FEC performance is highly sensitive to the interplay between redundancy, symbolization, and network conditions. No single static configuration performs well across all scenarios. This motivates TAROT, a fully adaptive FEC controller that formulates the selection of the optimal configuration $(n, k, S)$ for *every segment* as a constrained optimization problem based on real-time network and playback conditions.

### 3.2.1 Problem Formulation. 
The FEC parameter selection problem can be formally expressed as a Mixed-Integer Linear Program (MILP). While the problem can be formally cast as an MILP, the finite and manageable size of the candidate set $C$ (typically a few hundred configurations) makes a direct enumeration approach computationally feasible and equivalent to solving the MILP. This avoids the overhead of a generic MILP solver. Let the system state for each segment be characterized by the telemetry vector $\mathbf{s} = (br, bl, pl, gp)$, where $br$ is the requested playback bitrate (bps), $bl$ is the client buffer occupancy (seconds), $pl$ is the short-term packet-loss probability; and $gp$ is the measured link goodput (bps).

Let $C$ be the finite library of candidate FEC configurations, where each candidate $c \in C$ is defined by the tuple $(n_c, k_c, S_c, \text{codec}_c)$ with $n_c, k_c \in \mathbb{Z}^+$ and $S_c \in \mathbb{Z}^+$ (symbol size in bytes).

We introduce binary decision variables $x_c \in \{0, 1\}$ for each candidate $c \in C$, where $x_c = 1$ indicates configuration $c$ is selected. The MILP formulation is:

$$\begin{aligned}
\underset{x_c}{\text{minimize}} \quad & \sum_{c \in C} J(c) \cdot x_c \\
\text{subject to} \quad & \sum_{c \in C} x_c = 1 \quad \text{(Single selection)} \\
& \sum_{c \in C} \left( \frac{k_c}{n_c} - \alpha(bl, br) \cdot pl \right) x_c \geq 0 \quad \text{(Feasibility)} \\
& x_c \in \{0, 1\}, \quad \forall c \in C
\end{aligned} \quad (1)$$

where the objective function $J(c)$ for each candidate is:

$$J(c) = w_{\text{loss}} P_{\text{loss}}(c) + w_{\text{over}} P_{\text{over}}(c) + w_{\text{blk}} P_{\text{blk}}(c) \quad (2)$$

The protection margin $\alpha(bl, br)$ determines the minimum required redundancy ratio given the observed loss. Here, $B_{\text{eff}}$ denotes the effective client playback buffer level (in seconds), and $B_{\text{crit}}$ is a critical buffer threshold below which TAROT increases protection to avoid imminent rebuffering. It is defined as:

$$\alpha(bl, br) = \alpha_{\min} + \alpha_B \cdot \max(0, B_{\text{crit}} - B_{\text{eff}}) - \alpha_h \cdot \min\left(\frac{gp - br}{br}, h_{\text{cap}}\right) \quad (3)$$

where $\alpha_{\min} = 1.0$ is the baseline protection factor, $\alpha_B = 0.5$ increases the margin when buffer is critically low, and $\alpha_h = 0.5$ reduces the margin when abundant bandwidth headroom exists. This ensures stronger protection during buffer crises while allowing efficiency during favorable conditions. The adaptive weights $w_{\text{loss}}, w_{\text{over}}, w_{\text{blk}}$ are normalized ($\sum w_\cdot = 1$) and adjust dynamically based on $\mathbf{s}$: High $pl$ increases $w_{\text{loss}}$; low $gp$ or tight $bl$ increases $w_{\text{over}}$; and critically low $bl$ increases $w_{\text{blk}}$.

### 3.2.2 The TAROT Controller Architecture. 
While the MILP formulation provides theoretical grounding, TAROT employs efficient enumeration for practical deployment. As shown in Figure 3, TAROT consists of four modules that implement this optimization.

**Module 1: Telemetry Monitor.** For every segment request, TAROT collects the cross-layer state vector $\mathbf{s} = (br, bl, pl, gp)$. These signals determine both the minimum protection required to mask losses and the system's tolerance to overhead and delay.

**Module 2: Feasible Set Pruner.** This module applies the MILP feasibility constraint to prune the candidate library $C$. It calculates the minimum required redundancy fraction $r_{\min} = \alpha(bl, br) \cdot pl$ and constructs the feasible set $C_f$:

$$C_f = \left\{ c \in C \ \middle| \ \frac{k_c}{n_c} \geq r_{\min} \right\}$$

This implements the constraint satisfaction phase of the MILP, reducing the search space.

**Module 3: Scoring Engine.** Each candidate $c \in C_f$ is evaluated using the multi-objective penalty model from Equation 2.

**Loss Protection Penalty ($P_{\text{loss}}$):** Penalizes configurations where the repair budget is insufficient for the expected loss. Let $L_{\text{eff}} = n \cdot pl$ be the expected number of lost source symbols:

$$P_{\text{loss}}(c) = \max\left(0, \ L_{\text{eff}} - \beta \cdot k\right)^2 \quad (4)$$

where the codec efficiency factor $\beta \in [0, 1]$ accounts for the decoding performance of different FEC families. For Maximum Distance Separable (MDS) codes like Reed-Solomon, $\beta = 1.0$ since $k$ repair symbols can recover any $k$ lost symbols. Near-MDS codes like RaptorQ use $\beta = 0.99$, while simpler codes like XOR-based parity use $\beta < 1.0$ due to their limited recovery patterns. These values are derived from codec specifications and empirical validation. The quadratic form aggressively penalizes configurations that fall significantly short of required protection, making the optimization risk-averse to decoding failures. This ensures TAROT prioritizes reliable decoding over marginal bandwidth savings, as failed decoding triggers costly retransmissions that degrade QoE more severely than moderate overhead.

**Overhead Penalty ($P_{\text{over}}$)**: Penalizes bandwidth consumption from redundancy that exceeds the current allowance:

$$P_{\text{over}}(c) = \left(\max\left(0, \frac{k}{n} - \sigma_{\text{free}}(gp, br, bl)\right)\right)^{\alpha_{\text{over}}} \quad (5)$$

The overhead allowance function $\sigma_{\text{free}}(gp, br, bl)$ represents the maximum tolerable FEC overhead given current network conditions. It is computed as:

$$\sigma_{\text{free}} = \min\left(\sigma_{\text{cap}},\ \sigma_0 + \kappa_B \cdot \max(0, B_{\text{crit}} - B_{\text{eff}}) + \kappa_h \cdot \min\left(\frac{gp - br}{br}, h_{\text{cap}}\right)\right) \quad (6)$$

where $\sigma_0 = 0.01$ is the baseline overhead allowance, $\kappa_B = 0.02$ scales with buffer deficit (per second below $B_{\text{crit}}$), $\kappa_h = 0.03$ scales with available bandwidth headroom, and $\sigma_{\text{cap}} = 0.35$ imposes a hard ceiling. This formulation allows more FEC overhead when buffer is low (requiring protection) or when bandwidth headroom is abundant, while preventing excessive overhead that would degrade video quality. The convex penalty (with $\alpha_{\text{over}} = 1.5$) discourages excessive overhead that would degrade video quality, while the $\sigma_{\text{free}}$ allowance adapts to current network conditions—permitting more overhead during buffer crises or when bandwidth is abundant. This structure prevents FEC from consuming disproportionate bandwidth during stable periods.

**Blockization Penalty ($P_{\text{blk}}$)**: Penalizes timing risks. The total block delivery time is:

$$T_{\text{blk}} = \frac{(n+k)S}{gp} + T_{\text{enc}}(n, k, S, \text{codec})$$

$$P_{\text{blk}}(c) = \max\left(0, \frac{T_{\text{blk}} - \delta(bl)}{\delta(bl)}\right) \quad (7)$$

The deadline function $\delta(bl)$ defines the maximum acceptable block delivery time given current buffer conditions. It is computed as $\delta(bl) = \eta \cdot B_{\text{eff}}$, where $\eta = 0.5$ and $B_{\text{eff}}$ is the effective buffer level. This means the system tolerates block delivery times up to 50% of the current buffer duration, beyond which penalties apply. The penalty saturates at 1.0 when $T_{\text{blk}} \geq 2 \cdot \delta(bl)$.

The linear relative penalty beyond the $\delta(bl)$ deadline provides smooth degradation rather than hard constraints, allowing the optimizer to trade small timing violations against significant protection benefits when necessary. This flexibility is crucial for handling transient network conditions without causing abrupt quality switches.

*Module 4: FEC Configuration Selector.* This module solves the reduced MILP over the feasible set by selecting the configuration with minimum score:

$$c^* = (n^*, k^*, S^*) = \arg\min_{c \in C_f} J(c)$$

This yields the optimal redundancy rate $r^* = k^*/n^*$ for the segment.

To assess the robustness of our scoring model, we performed a small sensitivity analysis by varying two key hyperparameters: the buffer-awareness factor $\alpha_B$ and the loss-weight scaling $\lambda_p$—by ±20% around their calibrated values. Across all traces and both ABRs, average VMAF changed by only 1-2 units and rebuffering varied by less than 0.2%. This demonstrates that TAROT's performance is not the result of fragile hyperparameters. Instead, the multi-objective formulation itself provides structural resilience: its competing penalties (overhead, latency, loss-coverage, and deadline pressure) naturally balance one another, ensuring that small changes in individual weights do not meaningfully alter the selected FEC configuration or its QoE impact.

**Computational Complexity.** The enumeration approach performs three phases: (1) feasibility pruning ($O(|C|)$ comparisons), (2) scoring (15 operations per candidate), and (3) selection ($O(|C_f|)$ comparisons). Since $|C_f| \leq |C|$, the overall complexity is $O(|C|)$. With typical $|C| \approx 400$ and modern hardware (24-core CPU with 64 GB RAM), this yields $\approx 500\mu s$ decision overhead—negligible compared to 2-6s segment durations, ensuring real-time feasibility even on resource-constrained devices.

TAROT's MILP-based optimization balances loss protection, FEC overhead, and block delivery latency, avoiding the brittleness of static FEC settings while maintaining computational feasibility for real-time deployment. A comprehensive description of TAROT 's design details is provided in the appendix. Specifically, Appendix A presents the complete set of hyperparameters used by TAROT, along with their empirical calibration rationale and operational ranges. Appendix B contains all mathematical formulations and formal proofs, including correctness of the MILP equivalence, global optimality, constraint satisfaction guarantees, complexity analysis, Pareto optimality, robustness to measurement noise, and stability of the composite objective. These appendices collectively establish the theoretical soundness, optimality, and robustness properties of TAROT while keeping the main text focused on system design and experimental insights.

### 3.3 Real-World Deployment

Although our evaluation uses a controlled setup, TAROT is designed for integration into real-world streaming stacks built atop *unreliable or partially reliable* UDP-based transports such as QUIC [25], MPQUIC [57], RTP/SRT [48], and proprietary CDN transports [39]. These transports expose packet-delivery statistics to the application and rely on FEC rather than retransmissions, making them the natural deployment targets for TAROT. While reliable transports (e.g., TCP, SCTP) provide in-order recovery, TAROT remains compatible with application-layer FEC layers deployed above them (e.g., DASH over TCP [16]), as it requires only generic packet-delivery telemetry rather than transport-specific hooks.

In practice, TAROT operates on two counters per segment—packets sent and packets declared lost—which are available across nearly all media-delivery stacks. QUIC and MPQUIC export ACK ranges and loss reports via qlog [31]; RTP and SRT expose similar statistics; TCP/SCTP permit loss inference from SACK blocks or retransmission counters; and edge-based CDN reliability layers export per-chunk ACK/loss summaries. Cross-layer metrics such as client buffer level, playback bitrate, and request timing are exposed via CMCD [12] (Common Media Client Data) or via enhanced mechanisms such as the CAQ telemetry interface proposed in [50, 51]. These signals supply the $br$, $bl$, $pl$, and $gp$ inputs required by TAROT 's adaptive scoring model. Short-term loss per segment is computed as the ratio of lost to sent packets and smoothed using EWMA:

$$p = \lambda p_{\text{seg}} + (1 - \lambda)p_{\text{prev}},$$

with higher $\lambda$ for LLL mode and lower values for VoD.

The per-segment optimization in TAROT is intentionally lightweight: evaluating all $(n, k, S)$ configurations requires $\approx 500\ \mu s$, negligible relative to typical 2–6 s segment durations. Because TAROT depends only on standard packet-delivery telemetry and widely available cross-layer client signals, it integrates seamlessly into DASH/HLS/LLL players, CDN edge nodes performing on-the-fly FEC insertion, QUIC/WebRTC gateways, and UDP-based ultra-low-latency pipelines. Overall, TAROT aligns with the observability and control cadence of modern production streaming systems while remaining most effective in environments where retransmissions are costly or undesirable.

## 4 METHODOLOGY

### 4.1 SABRE-FEC

SABRE is a lightweight and widely-used simulator for evaluating HAS streaming logic under controlled network conditions. However, the original SABRE framework only models throughput and RTT variations and lacks support for packet-level loss or any form of Forward Error Correction. To accurately study the interaction between loss, redundancy, and adaptive video streaming, we extend SABRE into SABRE-FEC, shown in Figure 4. Our modifications (highlighted in grey) introduce two new components: a packet-loss simulation module and a modular FEC plugin. Together, these additions enable SABRE-FEC to reproduce realistic loss patterns and evaluate per-segment FEC decisions within the client's adaptive scheduling loop. In SABRE, the video manifest contains only the per-segment sizes for each representation rather than actual media chunks. SABRE-FEC uses these sizes as the payload input to compute symbolization, redundancy, and FEC-protected segment size.

*4.1.1 Packet-Loss Simulation Module.* Accurately modeling packet loss is essential for evaluating any FEC mechanism, yet loss is difficult to simulate because its effect is mediated through transport-layer dynamics. A single loss event can trigger congestion-window collapse, retransmission bursts, ACK compression, pacing disruptions, and interactions with delayed-ACK policies. Consequently, the throughput degradation caused by loss is rarely linear in the loss ratio, and simulators that simply scale bandwidth by $(1 - L)$ dramatically misestimate the impact of loss on streaming artifacts.

***Design Principle.*** If the loss ratio is $L$, then a fraction $L$ of transmitted packets is unavailable for contributing useful payload. However, the effective throughput penalty is shaped not only by the *volume* of data lost but also by how transport protocols react to loss. Loss-based congestion-control algorithms (e.g., CUBIC, Reno), QUIC recovery, and RTP/SRT pacing all exhibit a characteristic *superlinear* collapse under non-trivial loss [8, 22, 38, 40]. Empirical studies [4, 15, 38] repeatedly show that once loss exceeds 1–2%, the goodput degrades faster than $1 - L$ would suggest.

***Design Objective.*** SABRE-FEC requires a loss model that:
(1) captures the dominant *transport-level throughput collapse*,
(2) remains *protocol-agnostic* (no congestion-window modeling),
(3) is lightweight enough to run *once per segment*.

Full QUIC/TCP recovery simulation is too expensive, while a fully linear approximation is too inaccurate. We therefore adopt an analytically motivated throughput model inspired by the classical TCP loss models [2, 4, 5, 24] and recent loss-sensitivity analyses of QUIC and 5G RAN behavior [33, 45, 52].

**Analytical Model.** We define effective bandwidth under loss as:

$$B_{\text{loss}} = \frac{B_{\text{link}}}{1 + \gamma L \cdot 100 \cdot \sqrt{L}}, \quad where \quad (8)$$

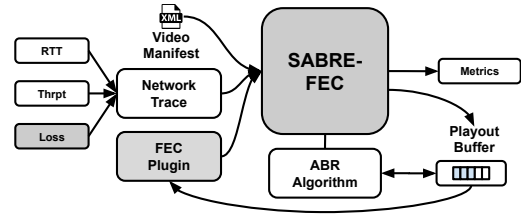

Figure 4: SABRE-FEC architecture.

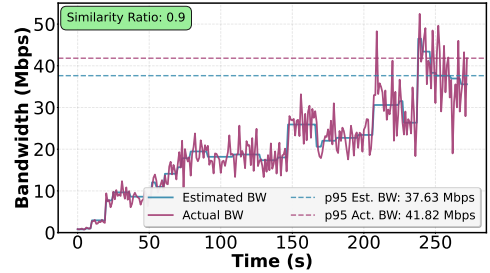

Figure 5: SABRE-FEC loss estimation accuracy.

- $B_{\text{link}}$ is the nominal link bandwidth from the trace,
- $L$ is the loss *ratio* (e.g., 1% loss $\Rightarrow L = 0.01$),
- $\gamma$ is a scaling parameter (default $\gamma = 0.5$).

Notably, $\gamma$ was selected empirically through calibration against real measurements obtained from the Vegvisir emulator, as shown in Figure 5. The $L\sqrt{L}$ term provides a superlinear collapse that mirrors established transport behavior:

- at very low loss, throughput roughly tracks $(1 - L)$,
- beyond 1–3% loss, goodput drops sharply reflecting TCP backoff,
- at high loss, the model approaches the nonlinear collapse seen in wireless links.

This model requires no ACK handling, no congestion window simulation, and no protocol instrumentation—making it suitable for large-scale per-segment experiments—while still capturing the qualitative shape of real transport behavior.

***Protocol-Agnostic Interpretation.*** Unlike transport-specific recovery models, Equation (8) applies unmodified to:

- QUIC and MPQUIC (via qlog-exposed loss),
- RTP/SRT/WebRTC transports with explicit loss counters,
- DASH/HLS over TCP with application-layer loss estimation

This makes loss simulation consistent across all evaluated schemes—No-FEC, static FEC, and TAROT. We note that while different congestion control algorithms (e.g., TCP CUBIC/Reno versus QUIC BBR) respond differently to packet loss, our model intentionally captures the dominant loss–goodput collapse trend using a single empirically calibrated scaling factor; protocol-specific calibration of $\gamma$ is left for future work.

To validate the accuracy of our loss–goodput formulation, we compared Eq. (8) against real measurements obtained from *Vegvisir* [19], an open-source network emulator that couples `tc netem` with containerized client/server endpoints. Vegvisir provides ground-truth packet delivery statistics under controlled impairments while preserving realistic TCP/QUIC congestion-control dynamics. We replayed more than 50 LTE and 5G traces [30, 41, 42, 56] and injected a controlled 5% packet-loss profile while running the throughput-based ABR. Figure 5 compares the *actual* goodput measured by Vegvisir client ("Observed Throughput") against the $B_{\text{loss}}$ predicted by Eq. (8). Across all network regimes, the analytical model remained within the **95th-percentile error envelope** of the simulator. Additionally, we compute a similarity ratio of 0.9, defined as the fraction of time samples for which the predicted goodput deviates from the observed goodput by less than the 95th-percentile error bound across all traces. This level of agreement shows that the model captures the dominant effect of loss on transport goodput while avoiding the cost of full protocol simulation. Thus, it is accurate enough to drive TAROT 's segment-level FEC decisions, yet lightweight enough to execute online within the tight control loop of adaptive streaming.

Static FEC and adaptive FEC behave very differently under non-linear throughput collapse. Accurate modeling of loss-induced goodput reduction is essential; otherwise:

- static FEC schemes appear artificially strong (because the true bandwidth penalty is underestimated), and
- adaptive schemes appear unstable (as the redundancy–buffer–goodput coupling is distorted).

Our analytical loss module ensures SABRE-FEC captures the correct qualitative response to loss, thus enabling fair and realistic evaluation of TAROT.

*4.1.2 Modular FEC Plugin.* SABRE-FEC includes a modular FEC plugin that enables evaluating different coding families under a unified interface. This module captures the two fundamental costs of forward error correction: (i) the *bandwidth cost* of transmitting repair symbols, and (ii) the *computational cost* of encoding and decoding. These costs interact directly with the ABR throughput estimator and therefore must be modeled explicitly.

To reflect these interactions faithfully, the FEC layer computes an adjusted segment delivery time that incorporates both the inflation in transmitted bytes due to redundancy and the additional time required to generate repair symbols. Encoding introduces a delay before a segment becomes available for transmission, and this delay is explicitly added to the segment's download time.

On the receiver side, we make the standard assumption—consistent with modern DASH, HLS, and LL-HLS players—that decoding is performed on a non-blocking background thread. In such architectures, decoding latency does *not* contribute to measured throughput or the ABR's estimate of segment-fetch time; it only affects the rendering pipeline. In our experiments, decoding latency for RaptorQ, Reed–Solomon, and XOR is on the order of a few milliseconds and remains far below typical buffer levels (6 seconds for live-streaming and 60 seconds for on-demand streaming), making its impact on playout negligible. Under this assumption, only encoding latency lies on the critical path and must be included in throughput model.

We integrate three widely used FEC families into SABRE-FEC:

- **XOR (block parity):** minimal encoding cost; supports only single-loss recovery.
- **Reed–Solomon (RS):** MDS code with strong recovery guarantees but higher encoding cost.
- **RaptorQ (RQ):** fountain code supporting flexible block sizing and moderate computational cost.

TAROT's adaptive FEC adjustment (as shown in Figure 3) is also implemented within this module. To parameterize encoding latency realistically, we benchmarked the open-source `quic-go-fec` implementation [32, 36] and measured average per-block encoding time across different $(n, k, S)$ settings. Notably, our measurements show that RaptorQ incurs an average encoding cost of approximately 22 ns per byte, while Reed–Solomon requires about 35 ns per byte. These values are injected into SABRE–FEC to model realistic computational overhead and ensure that the impact of encoding latency is reflected in the segment fetch time to mimic real-world behavior.

***Bandwidth Reduction Under FEC Overhead.*** Let $n$ denote the number of source symbols, $k$ the number of repair symbols, and $S$ the symbol size. The bandwidth inflation introduced by FEC is captured by the overhead factor:

$$\text{overhead\_factor} = \frac{n+k}{n}.$$

The FEC module adjusts the *effective* bandwidth exposed to the ABR logic by accounting for both (i) transmitted redundancy and (ii) the extent to which repair symbols mitigate packet loss. The resulting payload goodput is:

$$B_{\text{FEC}} = \frac{B_{\text{link}}(1 - l_{\text{eff}})}{\text{overhead\_factor}}, \qquad (9)$$

where $l_{\text{eff}}$ is the *residual loss* after applying the code's recovery capability and is computed as:

$$l_{\text{eff}} = \begin{cases} L\left(0.4 + 0.6\left(1 - \frac{\text{headroom}}{\text{coverage}}\right)\right), & L \leq \text{coverage}, \\ L - 0.8\,\text{coverage}, & L > \text{coverage}, \end{cases}$$

clipped to $[0, L]$, where $L$ is the observed loss ratio and coverage $= k/(n+k)$. This model captures the diminishing returns of redundancy: when coverage comfortably exceeds the loss rate, FEC suppresses most losses; when coverage is insufficient, only a fraction is mitigated. Equation (9) therefore represents the payload goodput visible to the ABR algorithm. Combined with encoding latency, SABRE-FEC models both bandwidth inflation and computational overhead in a manner aligned with operational streaming systems.

As the FEC plugin exposes a uniform API for computing FEC-protected throughput, residual loss, and encoding latency, TAROT can be integrated with *any* FEC scheme—XOR, RS, or RQ—without modification. As shown in Section 4.3, TAROT consistently improves QoE across all three code families by dynamically selecting $(n, k, S)$ to balance redundancy, blockization delay, and computational cost.

### 4.2 Simulation Setup

*4.2.1 Video Contents and Parameters.* For all online evaluations of TAROT, we use the *NVIDIA Asteroid* (NVA) video dataset [37], encoded in AVC [58] with ten bitrate representations ranging from 288p@0.5 Mbps to 2160p@40 Mbps (4K). Experiments are conducted in two streaming scenarios: *Video-on-Demand* (VoD), using 4s segments and a maximum playback buffer of 60s, and *Low-Latency Live* (LLL), using 2s segments and a 6s playback buffer. Each experiment streams the full 4 min 30 s duration of the content, ensuring that long-term adaptation behavior is captured and that steady-state performance dominates transient startup effects.

*4.2.2 Network Traces.* For the online evaluation of TAROT, we employ four widely used real-world network traces: the *Netflix 5G* trace [41], the *LTE Belgium* trace [56], the *Amazon 5G* trace [41], and the *Cascade* trace [20] to simulate realistic 5G↔4G handover dynamics. The characteristics of these traces are summarized in Table 1. Each trace is evaluated under four loss profiles: (i) no loss, (ii) a consistent 1% loss rate, (iii) a consistent 5% loss rate, and (iv) a variable loss model drawn from a random sampler between 0–5%, capturing the volatility observed in congested cellular networks.

Table 1: Network Trace Data.

| Network Trace | Avg. Thrpt. (Mbps) | Std. Dev. (Mbps) |
|---|---|---|
| Netflix5G [41] | 33 | 18 |
| Amazon5G [41] | 25 | 10 |
| LTE Belgium [56] | 20 | 5 |
| Cascade [20] | 30 | 15 |

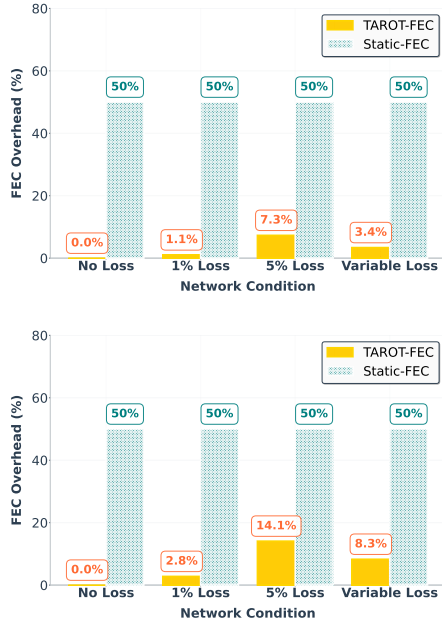

Figure 6: Static FEC vs TAROT-FEC overhead comparison averaged over various network traces and ABRs for LLL mode (top) and VoD mode (bottom).

*4.2.3 ABR Algorithms.* We evaluate TAROT using the two ABR schemes provided by the original SABRE [53] framework: (i) a *throughput-based* (THR) ABR that selects the highest sustainable bitrate based on recent download rates, and (ii) the *Dynamic* [55] ABR, which combines throughput estimation with BOLA-style buffer-aware decisions as described in the SABRE simulator. In the VoD configuration, both ABRs operate normally because the playback buffer can grow up to 60s, allowing the Dynamic ABR to exploit its buffer-driven logic. However, in the LLL setup, the buffer is capped at only 6s. Under this tight buffer constraint, the Dynamic ABR repeatedly collapses to the lowest quality level due to its BOLA-based utility function. As a result, for LLL we report results only for the THR ABR, whereas both ABRs are evaluated in the VoD mode.

*4.2.4 FEC Strategies.* To rigorously evaluate TAROT across different coding families, we consider all FEC mechanisms integrated into our SABRE-FEC framework: (1) **No FEC**, (2) **RaptorQ (RQ)**, (3) **Reed–Solomon (RS)**, (4) **TAROT-enhanced RaptorQ (RQ*)**, and (5) **TAROT-enhanced Reed–Solomon (RS*)**.

We exclude learning-based FEC schemes (e.g., RL-FEC, RL-AFEC) because these methods focus only on tuning redundancy ratios and require pre-training, whereas TAROT's optimization-driven design jointly selects the full $(n, k, S)$ configuration and operates online in SABRE-FEC without any prior training, making such comparisons inequitable. More broadly, learning-based FEC schemes operate as black-box controllers with limited interpretability and no explicit fallback guarantees, whereas TAROT's white-box optimization explicitly exposes and balances the trade-offs between loss protection, overhead, and timing constraints. Both RQ and RS use the same default FEC parameters ($n=20, k=10, S=64$) taken from the publicly available FEC implementations [32, 35], which we integrated into SABRE-FEC. We intentionally exclude XOR-based FEC from our results. XOR operates with a fixed 2-source/1-parity configuration that (i) does not expose $(n, k, S)$ as tunable variables, (ii) imposes a rigid 50% overhead, and (iii) provides only single-symbol recovery. This makes XOR structurally incompatible with TAROT's adaptive parameter selection, since it offers no meaningful degrees of freedom for optimization. Furthermore, initial experiments showed that XOR consistently underperforms compared to RQ and RS across all traces and loss levels due to its poor recovery strength and high overhead. Excluding XOR therefore leads to clearer, fairer comparisons among FEC schemes that can actually benefit from adaptive optimization.

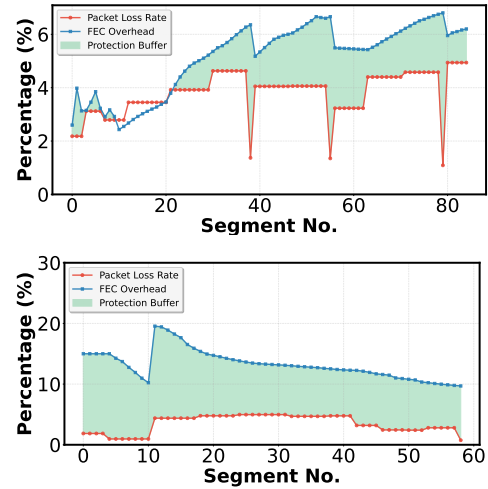

Figure 7: TAROT dynamic per-segment FEC adjustment for LLL (top) and VoD (bottom). "Protection Buffer" denotes the redundancy selected by TAROT per segment, expressed as the fraction of protection symbols in the FEC block ($r = k/n$).

*4.2.5 Performance Metrics.* We evaluate TAROT using widely adopted performance metrics: (1) Video Multi-Method Assessment Fusion (VMAF) [43], (2) Rebuffering Duration (RD), and (3) FEC overhead, measured as the fraction of additional repair bytes transmitted relative to the original payload size.

*4.2.6 System Configurations.* All experiments ran on Ubuntu 22.04.3 with a 24-core CPU, 64 GB RAM, and Python 3.11.

### 4.3 Results and Discussions

*4.3.1 Impact of TAROT on FEC Overhead.* As shown in Figure 6, TAROT-FEC completely suppresses redundancy when no loss is observed, in both VoD and LLL settings. This behavior arises from TAROT's design: FEC is activated *only* when the loss estimator detects non-zero packet loss, making the mechanism inherently reactive rather than always-on. In contrast, static FEC continues to inject redundancy even when the path is loss-free, incurring unnecessary overhead. Once loss is introduced—whether as a constant 1% or 5% rate, or under a variable loss profile sampling between 0% and 5%—TAROT reduces overhead by up to 46% in LLL mode and 41% in VoD mode. This reduction stems from its multi-objective optimization, which selects the smallest $(n, k, S)$ configuration sufficient to meet the recovery budget imposed by the observed loss. Figure 7 illustrates this adaptive behavior for a 5% loss scenario on the Amazon 5G trace under the throughput-based ABR. Over the duration of the session, TAROT reacts promptly to fluctuations in loss, maintains adequate recovery coverage, and avoids the excessive redundancy produced by static FEC.

Table 2: Quality–Overhead tradeoff averaged over all traces and ABRs for LLL and VoD modes. Here, $r$ denotes FEC overhead.

| Mode | Loss | No Loss | | | 1% Loss | | | 5% Loss | | | Variable Loss | | |
|---|---|---|---|---|---|---|---|---|---|---|---|---|---|
| | FEC type | VMAF | RD | $r$ (%) | VMAF | RD | $r$ (%) | VMAF | RD | $r$ (%) | VMAF | RD | $r$ (%) |
| LLL | No FEC | **60.61** | 1.51 | 0 | **59.58** | 1.71 | 0 | 50.05 | 1.26 | 0 | 55.05 | 1.49 | 0 |
| | RQ | 50.65 | 1.05 | 50 | 50.11 | 0.88 | 50 | 50.27 | 1 | 50 | 49.86 | 0.78 | 50 |
| | RS | 50.54 | 0.78 | 50 | 50.45 | 1.29 | 50 | 49.75 | 1.2 | 50 | 49.34 | 1.18 | 50 |
| | RQ* | **60.95** | 1.5 | 0 | **60.14** | 1.42 | 1 | **59.71** | 1.22 | 7.05 | **62.77** | 1.36 | 3.4 |
| | RS* | **60.92** | 1.5 | 0 | **59.92** | 1.84 | 1.2 | **59.71** | 1.45 | 7.55 | **62.60** | 1.37 | 3.4 |
| VoD | No FEC | 78.97 | 0.35 | 0 | 75.98 | 0.12 | 0 | 62.13 | 0.08 | 0 | 70.93 | 0.14 | 0 |
| | RQ | 65.46 | 0.03 | 50 | 64.15 | 0.04 | 50 | 62.80 | 0.07 | 50 | 62.63 | 0.04 | 50 |
| | RS | 62.52 | 0.04 | 50 | 63.17 | 0.04 | 50 | 62.68 | 0.08 | 50 | 64.05 | 0.05 | 50 |
| | RQ* | **79.50** | 0.37 | 0 | **78.33** | 0.14 | 2.75 | **71.90** | 0.07 | 14.2 | **76.03** | 0.01 | 8.1 |
| | RS* | **79.35** | 0.13 | 0 | **79.84** | 0.19 | 2.85 | **75.69** | 0.08 | 14 | **78.04** | 0.02 | 8.5 |

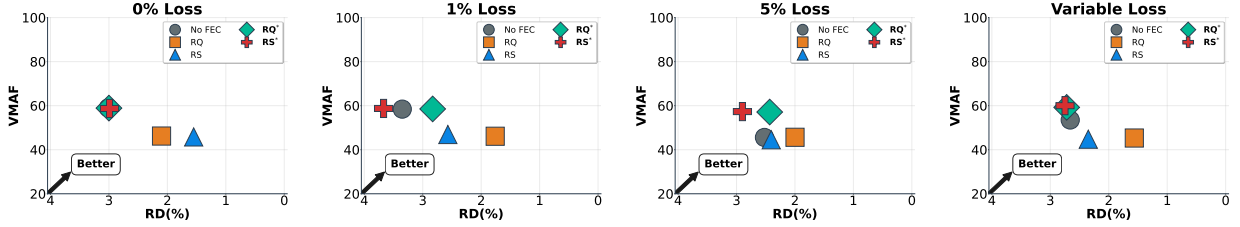

Figure 8: Impact of FEC on VMAF vs RD analysis for Cascade for ABR:THR and mode:LLL.

These overhead savings translate directly into QoE improvements: with fewer repair bytes, more of the link bandwidth remains available for video segments, enabling the ABR to select higher-quality representations in both LLL and VoD modes (as further detailed in Section 4.3.2). Notably, the savings are more pronounced in VoD due to its longer segment durations, which otherwise amplify the cost of static redundancy.

4.3.2 *Impact of TAROT on HAS Video Streaming.*
▷**LLL mode.** Figure 8 illustrates the comparison between No FEC, the static RQ and RS baselines, and their TAROT-enhanced variants (RQ* and RS*) for the *Cascade* trace under THR ABR in LLL mode across different loss profiles. Under the no-loss scenario, as the network experiences *no* packet loss, TAROT is never triggered for RQ* and RS*, and these adaptive variants behave similar to No FEC.

In contrast, the static RQ and RS schemes are locked into their fixed ($n$=20, $k$=10, $S$=64) configuration, which imposes a 50% overhead even when no protection is required. This unconditional overhead suppresses the effective bandwidth exposed to the ABR logic and forces the player to choose lower-quality representations throughout the session. As a result, static RQ and RS trail behind No FEC and the TAROT-enhanced variants by 13 VMAF units (roughly 2 JNDs). Their slightly lower rebuffering ratio (about 1%) is simply a consequence of streaming at consistently lower bitrates; this minor reduction is far from sufficient to offset the substantial perceptual quality loss caused by the inflated overhead.

Under the 1% loss scenario, we observe a similar trend. Because the loss level remains mild, No FEC, RQ*, and RS* achieve nearly indistinguishable VMAF, while the static RQ and RS baselines continue to lag behind by roughly 12 VMAF units (2 JNDs) due to their rigid 50% overhead. TAROT does react to the measured loss and activates its estimator, but the required protection at 1% loss is far smaller than 50%. Consequently, TAROT selects lightweight FEC configurations whose overhead remains close to zero, allowing it to preserve almost the same effective throughput and representation choices as No FEC. In contrast, No FEC cannot mask packet losses through FEC, and the end-to-end retransmission delay at 1% loss is comparable to the combined cost of mild FEC overhead and encoding latency. As a result, No FEC, RQ*, and RS* achieve essentially the same QoE tradeoffs in this regime, and all schemes exhibit similar rebuffering ratios.

Under the 5% loss scenario, RQ* and RS* deliver a clear advantage by achieving 12 VMAF units (2 JNDs) higher than No FEC, RS and RQ. This improvement arises because TAROT identifies the minimum overhead needed to withstand the elevated loss rate, applying just enough protection to prevent retransmission-induced stalls while preserving most of the available bandwidth. In contrast, No FEC degrades sharply at 5% loss, as retransmissions become frequent and increasingly costly, leading to substantial quality drops. Surprisingly, the static RQ and RS schemes fail to outperform No FEC and perform comparably even in this high-loss regime. Although FEC should offer clear benefits at 5% loss, their fixed 50% overhead suppresses effective throughput so severely that both static schemes continue to lag behind TAROT-enhanced variants by nearly 12 VMAF units, despite transmitting significantly more redundancy.

Under the variable-loss scenario, where the packet loss fluctuates between 0% and 5%, we observe a similar but slightly narrower performance gap compared to the fixed 5% loss condition. Despite the variability, RQ* and RS* still achieve a consistent improvement of 6 VMAF units (1 JND) over No FEC. This gain again reflects

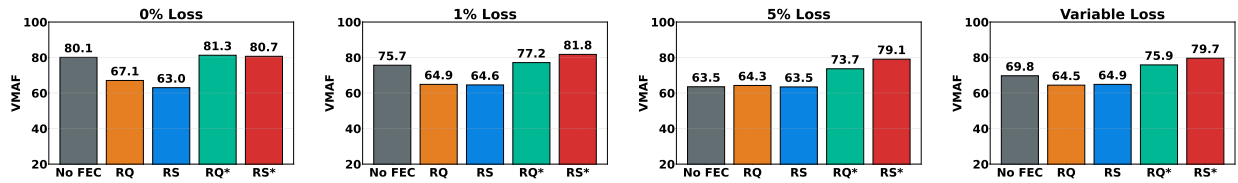

Figure 9: Impact of FEC on VMAF vs RD analysis for Netflix 5G for ABR:THR and mode:VoD.

TAROT's ability to adjust its overhead dynamically to match the instantaneous loss level, applying protection only when needed. By contrast, the static RQ and RS schemes continue to underperform: they fall behind No FEC by approximately 8 VMAF units (1 JND). As before, their fixed 50% overhead suppresses effective throughput across the entire session, even during intervals of low loss where such a large overhead provides no benefit. Consequently, both static FEC schemes remain unable to capitalize on the periods of reduced loss and continue to deliver lower perceptual quality than No FEC and the TAROT-enhanced variants.

Notably, the remaining network traces exhibited the same qualitative behavior as the Cascade trace. For brevity, we report the cross-trace averages for LLL mode in Table 2, comparing No FEC, static FEC (RS, RQ), and our TAROT-enhanced variants (RQ* and RS*). The aggregated results reinforce three key findings: (i) TAROT's zero-overhead advantage in stable, loss-free conditions, (ii) its ability to balance protection and throughput under moderate loss with only minimal redundancy, and (iii) its superior resilience in high-loss scenarios, where adaptive FEC materially improves VMAF while using an order of magnitude less overhead than static schemes.

▷**VoD mode.** Figure 9 compares No FEC, the static RQ and RS baselines, and their TAROT-enhanced variants for the *Netflix 5G* trace under THR ABR across all loss profiles. Unlike LLL mode, VoD exhibited *zero* rebuffering events for all schemes and all loss conditions due to its substantially larger playback buffer (60s). Accordingly, we report only the VMAF results. The trends in VoD mode mirror those observed in LLL mode, but with *exaggerated* VMAF differences. Across multiple loss profiles, static RQ and RS lag behind No FEC and the TAROT-enhanced variants by more than 12 VMAF units (approximately 2 JNDs)—a gap amplified by VoD's higher operating bitrates. Because VoD maintains a deep buffer, the ABR algorithm quickly stabilizes at high-quality representations, making the *FEC overhead the dominant factor*: the rigid 50% redundancy of static RQ/RS suppresses effective throughput far more than in LLL mode and severely degrades VMAF.

The cross-trace averages in Table 2 for VoD confirm the same pattern observed in the Netflix 5G trace: TAROT consistently preserves high VMAF in stable and moderate-loss regimes, and delivers markedly superior robustness under high loss, all while using an order of magnitude less overhead than static FEC. These findings underscore the importance of TAROT's adaptive per-segment FEC parameter selection. TAROT delivers the same core benefit in both LLL and VoD modes: zero cost in low loss, minimal overhead in moderate loss, and superior protection in high loss; which directly improves QoE, as evidenced by the higher VMAF scores.

*4.3.3 Why Redundancy Alone Is Not Enough?* Figure 10 compares TAROT-FEC with an R-FEC controller that adjusts only the redundancy ratio while keeping $(n, S)$ fixed. Although R-FEC outperforms static FEC (Table 2) in LLL mode, it still trails TAROT-FEC by roughly 6 VMAF units (1 JND). In VoD, the gap narrows but remains perceptually meaningful: while TAROT-FEC delivers nearly 2 JND of improvement over static FEC, R-FEC recovers only about 1 JND. The reason is structural— redundancy alone cannot satisfy the timing and recovery constraints imposed by the FEC block configuration. R-FEC (Figure 10) uses on average 18% more redundancy than TAROT-FEC across both modes yet achieves lower perceptual quality. Effective loss protection depends jointly on $(n, k, S)$, which determine encoding latency, repair capability, and per-block transmission cost. TAROT adapts this full configuration, not just redundancy, enabling consistently higher QoE across network conditions.

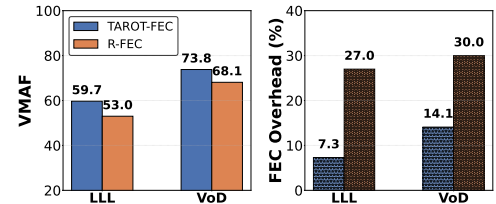

Figure 10: Ablation study: TAROT vs. a redundancy-only controller (R-FEC), averaged across all traces, ABRs, and FEC schemes at 5% packet loss.

## 5 LIMITATIONS AND FUTURE WORK

TAROT, while effective across diverse traces and ABRs, has several limitations. First, our loss–goodput formulation captures the dominant superlinear collapse but abstracts away transport-specific behaviors (congestion-window evolution, ACK timing, pacing artifacts). Although SABRE-FEC provides controlled and repeatable conditions, it remains a simulation framework and while our variable-loss model serves as a first-order approximation of network volatility: it captures short-term fluctuations but does not explicitly represent correlated burst losses common in mobile, Wi-Fi, and satellite links. Extending TAROT to bursty-loss processes (Gilbert–Elliott [34]) and validating its behavior under full QUIC/TCP emulation or hardware testbeds represents an important direction for future work.

Second, encoding latency is modeled using software-based RaptorQ and Reed–Solomon implementations. Hardware-accelerated or SIMD-optimized encoders may shift the latency–overhead trade-off, suggesting a need for device-specific profiling and adaptive complexity control. Similarly, TAROT currently adapts at segment granularity; ultra-low-latency workloads (e.g., sub-second live gaming or WebRTC pipelines) may benefit from chunk- or frame-level FEC decisions. Third, TAROT is intentionally content-agnostic: it applies the same protection strategy to all segments without analyzing codec structure or frame importance. This choice keeps the controller lightweight, avoids video parsing at the edge, and ensures applicability across codecs and delivery formats. However, it is inherently suboptimal from a perceptual standpoint—losing an I-frame can disrupt an entire GOP, whereas errors in P/B-frames have far smaller impact. Without distinguishing such differences, TAROT may over-protect less critical data and under-protect the most visually significant segments. Incorporating content-aware signals (e.g., frame type, motion complexity, saliency) is a promising direction for improving perceptual efficiency.

Fourth, while TAROT demonstrates strong performance through optimization-driven adaptation, we intentionally avoided machine learning components to ensure provable optimality guarantees, minimal computational overhead ($\approx$500$\mu$s), and immediate effectiveness without training data. However, future work could explore hybrid approaches where TAROT's optimization core is enhanced with learned components—for instance, using LSTM networks to predict loss patterns or reinforcement learning to adapt weight parameters based on long-term QoE objectives.

Finally, our evaluation focuses on LTE/5G mobile traces. Broader conditions such as Wi-Fi 6/7 contention, mmWave mobility, satellite links, or backbone congestion may exhibit different temporal loss structures. Extending TAROT to these environments, and exploring deeper cross-layer cooperation (e.g., AQM, RAN queueing signals), represents a promising direction for future deployment.

## 6 CONCLUSION

We introduced TAROT, an optimization-driven FEC controller that selects redundancy, symbolization, and block size per segment using a lightweight multi-objective scoring model. TAROT adapts to real-time loss, goodput, and buffer conditions and avoids the rigidity of static FEC schemes. To support realistic evaluation, we developed SABRE-FEC, a high-fidelity extension of SABRE that incorporates a nonlinear loss–goodput model and a modular FEC layer with measured encoding costs. Across LTE, 5G, and handover traces in both LLL and VoD settings, TAROT consistently reduced overhead (up to 45%), improved perceptual quality (+10% over No FEC and +20% over static FEC), and lowered rebuffering, all while imposing only $\approx$ 500 $\mu$s per-segment decision cost. TAROT relies solely on standard packet-delivery telemetry, and thus can be integrated into DASH/HLS clients, CDN edges, or QUIC-based delivery stacks without protocol modifications. Beyond enabling TAROT, our SABRE-FEC framework serves as a reusable benchmarking tool for evaluating future FEC schemes: it provides a controlled, high-fidelity playground for rapid testing of redundancy strategies, code families, and parameter tuning logic under realistic loss patterns and ABR dynamics. Overall, TAROT offers a practical and effective path toward adaptive, optimization-driven FEC parameter tuning for modern video streaming, enabling higher QoE and more efficient bandwidth usage across volatile network conditions.

**Acknowledgments**. This research is funded by the Natural Sciences & Engineering Research Council (NSERC)-Discovery Grant RGPIN-2023-04744 and NSERC Alliance (ALLRP 586434-23)-Mitacs Accelerate (IT36792) grant.

## A TAROT MODEL HYPERPARAMETERS

Table 3 lists all the hyperparameters used in TAROT adaptive FEC adjustment. All hyperparameters were empirically calibrated on a validation set comprising 10% of available network traces. Buffer thresholds ($B_{\text{sat}}$ = 6.0s, $B_{\text{crit}}$ = 3.0s) align with typical segment durations and critical playback limits. Protection factors ($\alpha_{\min}$ = 1.0, $\alpha_B$ = 0.5, $\alpha_h$ = 0.5) ensure 1:1 baseline protection while adapting to buffer and bandwidth conditions. Overhead parameters ($o_0$ = 0.01, $k_B$ = 0.02, $k_h$ = 0.03) permit minimal baseline overhead with controlled increases during adverse conditions. Weight adaptation parameters balance loss protection, overhead control, and timing constraints across diverse network scenarios. Algorithm 1 summarizes the overall flow of TAROT adaptive per-segment FEC adjustment.

## B MATHEMATICAL PROOFS

### B.1 Proof of MILP Formulation Correctness

THEOREM B.1 (MILP EQUIVALENCE). *The enumeration approach in Algorithm 1 solves the same optimization problem as the MILP formulation in Equation (1).*

PROOF. Let $C_f$ be the feasible set after constraint pruning. The MILP objective is:

$$\min \sum_{c \in C} J(c) x_c \quad \text{subject to} \quad \sum_{c \in C} x_c = 1, \quad x_c \in \{0, 1\}$$

Since exactly one $x_c$ = 1 and all others are 0, the objective reduces to $J(c)$ for the selected configuration. Algorithm 1 evaluates $J(c)$ for all $c \in C_f$ and selects arg min, which is equivalent to solving the MILP over the pruned feasible set. □

**Table 3: TAROT Hyperparameter Values.**

| Parameter | Value |
|---|---|
| *Buffer Management* | |
| $B_{\text{sat}}$ | 6.0 s |
| $B_{\text{crit}}$ | 3.0 s |
| $h_{\text{cap}}$ | 2.0 |
| *Loss Protection* | |
| $\alpha_{\min}$ | 1.0 |
| $\alpha_B$ | 0.5 |
| $\alpha_h$ | 0.5 |
| *Overhead Control* | |
| $o_0$ | 0.01 |
| $k_B$ | 0.02 |
| $k_h$ | 0.03 |
| $o_{\text{cap}}$ | 0.35 |
| $\alpha_{\text{over}}$ | 1.5 |
| *Blockization Control* | |
| $\eta$ | 0.5 |
| hardcap$_{\text{tblk}}$ | 1.5 |
| *Weight Adaptation* | |
| $w_{\text{loss}}^{\min}$ | 0.5 |
| $\lambda_p$ | 6.0 |
| $p_{\text{cap}}$ | 0.15 |
| $w_{\text{over}}^{\min}$ | 0.5 |
| $\lambda_B$ | 0.5 |
| $\lambda_h$ | 0.4 |
| $w_{\text{blk}}^{\min}$ | 0.3 |
| $\lambda_{\text{risk}}$ | 0.6 |
| $\lambda_{hneg}$ | 0.6 |

The constraint $\sum_{c \in C}(\frac{k_c}{n_c} - \alpha \cdot pl)x_c \geq 0$ is enforced by the feasibility pruning, ensuring only configurations satisfying $\frac{k_c}{n_c} \geq \alpha \cdot pl$ are considered. □

**Algorithm 1** TAROT: Optimal FEC Parameter Selection per Segment.

**Require:** loss $pl$, buffer $bl$, goodput $gp$, playback bitrate $br$, payload size $P$, candidate set $C = \{(n, S, r, \text{codec})\}$
1: $C_{\text{feasible}} \leftarrow \emptyset$
2: $B_{\text{eff}} \leftarrow \min(\max(bl, 0), B_{\text{sat}})$
3: $h_{\text{raw}} \leftarrow (gp - br)/\max(br, \epsilon)$
4: $h \leftarrow \min(\max(h_{\text{raw}}, -10.0), 10.0)$
5: $h_+ \leftarrow \max(0, h), h_- \leftarrow \max(0, -h)$
　　　　　　　　　　▷ Compute protection margin $\alpha(bl, br)$
6: $\alpha \leftarrow \alpha_{\min} + \alpha_B \cdot \max(0, B_{\text{crit}} - B_{\text{eff}}) - \alpha_h \cdot \min(h_+, h_{\text{cap}})$
7: $\alpha \leftarrow \max(0.5, \alpha)$
8: **for** each $(n, S, r, \text{codec})$ in $C$ **do**
9: 　　$k \leftarrow \lceil rn \rceil, T \leftarrow n + k, o \leftarrow k/n, cov \leftarrow k/T$
　　　　　　　　　　　　　　　　　　　　▷ Feasibility check
10: 　　**if** $o < \alpha \cdot pl$ **then continue**
11: 　　**end if**
　　　　　　　　　　　　　　▷ Loss effectiveness $f_{\text{loss}}(pl, cov)$
12: 　　**if** $pl \leq cov$ **and** $cov > 0$ **then**
13: 　　　　$headroom \leftarrow cov - pl$
14: 　　　　$reduction \leftarrow headroom/cov$
15: 　　　　$l_{\text{eff}} \leftarrow pl \cdot (0.4 + 0.6 \cdot (1 - reduction))$
16: 　　**else**
17: 　　　　$l_{\text{eff}} \leftarrow pl - 0.8 \cdot cov$
18: 　　**end if**
19: 　　$l_{\text{eff}} \leftarrow \max(0, \min(l_{\text{eff}}, pl))$
　　　　　　　　　　　▷ Recalculate headroom with FEC effects
20: 　　$G_{\text{payload}} \leftarrow gp \cdot (1 - l_{\text{eff}})/((1 - pl) \cdot (1 + o))$
21: 　　$h \leftarrow (G_{\text{payload}} - br)/\max(br, \epsilon)$
22: 　　$h \leftarrow \min(\max(h, -10.0), 10.0)$
23: 　　$h_+ \leftarrow \max(0, h), h_- \leftarrow \max(0, -h)$
　　　　　　　　　　　　　　　　　　▷ Overhead penalty
24: 　　$o_{\text{free}} \leftarrow o_0 + k_B \cdot \max(0, B_{\text{crit}} - B_{\text{eff}}) + k_h \cdot \min(h_+, h_{\text{cap}})$
25: 　　$o_{\text{free}} \leftarrow \max(0, \min(o_{\text{free}}, o_{\text{cap}}))$
26: 　　$P_{\text{over}} \leftarrow \max(0, o - o_{\text{free}})^{\alpha_{\text{over}}}$
　　　　　　　　　　　　　　　　　▷ Blockization penalty
27: 　　$t_{\text{blk}} \leftarrow (8 \cdot T \cdot S)/\max(gp, \epsilon)$
28: 　　**if** $t_{\text{blk}} > \text{hardcap}_{\text{tblk}} \cdot B_{\text{eff}}$ **then continue**
29: 　　**end if**
30: 　　$P_{\text{blk}} \leftarrow \min(1.0, \max(0, t_{\text{blk}}/(\eta \cdot B_{\text{eff}}) - 1))$
　　　　　　　　　　　▷ Loss penalty with codec efficiency
31: 　　$\beta \leftarrow \text{getCodecEfficiency}(\text{codec})$
32: 　　$P_{\text{loss}} \leftarrow \max(0, n \cdot pl - \beta \cdot k)^2$
　　　　　　　　　　　　　　　　　　　▷ Adaptive weights
33: 　　$w_{\text{loss}} \leftarrow w_{\text{loss}}^{\min} + \lambda_p \cdot \min(pl, p_{\text{cap}})$
34: 　　$w_{\text{over}} \leftarrow w_{\text{over}}^{\min} + \lambda_B \cdot (B_{\text{eff}}/B_{\text{sat}}) + \lambda_h \cdot \min(h_+, h_{\text{cap}})$
35: 　　$w_{\text{blk}} \leftarrow w_{\text{blk}}^{\min} + \lambda_{\text{risk}} \cdot \max(0, 1 - B_{\text{eff}}/B_{\text{crit}}) + \lambda_{hneg} \cdot h_-$
　　　　　　　　　　　▷ Normalize and compute final score
36: 　　$sum \leftarrow w_{\text{loss}} + w_{\text{over}} + w_{\text{blk}}$
37: 　　$w_{\text{loss}} \leftarrow w_{\text{loss}}/sum, w_{\text{over}} \leftarrow w_{\text{over}}/sum, w_{\text{blk}} \leftarrow w_{\text{blk}}/sum$
38: 　　$J \leftarrow w_{\text{loss}} \cdot P_{\text{loss}} + w_{\text{over}} \cdot P_{\text{over}} + w_{\text{blk}} \cdot P_{\text{blk}}$
39: 　　Add $(n, k, S, J)$ to $C_{\text{feasible}}$
40: **end for**
41: **return** $\arg\min_{(n,k,S,J) \in C_{\text{feasible}}} J$

## B.2 Proof of Optimality

THEOREM B.2 (GLOBAL OPTIMALITY). *For any finite candidate set $C$ and system state $\mathbf{s}$, Algorithm 1 returns the globally optimal configuration $c^*$ that minimizes $J(c)$ subject to the feasibility constraints.*

PROOF. Let $C_f = \{c \in C \mid \frac{k_c}{n_c} \geq \alpha \cdot pl\}$ be the feasible set.
1. **Feasibility**: By construction, $c^* \in C_f$, so $c^*$ satisfies all constraints.
2. **Optimality**: Algorithm 1 performs exhaustive enumeration over $C_f$, computing:
$$c^* = \arg\min_{c \in C_f} J(c)$$
Since $C_f$ is finite, the minimum exists and is found by enumeration.
3. **Global Scope**: Any $c \notin C_f$ is infeasible and cannot be optimal. Therefore, $c^*$ is globally optimal over $C$. □

## B.3 Proof of Constraint Satisfaction

THEOREM B.3 (DECODING GUARANTEE). *Under i.i.d. packet loss with probability $pl$, any configuration selected by TAROT guarantees successful decoding with probability at least $1 - \frac{1}{\alpha}$.*

PROOF. Let $X$ be the number of lost source symbols in a block of size $n$. Under i.i.d. loss:
$$\mathbb{E}[X] = n \cdot pl$$
The feasibility constraint requires:
$$\frac{k}{n} \geq \alpha \cdot pl \Rightarrow k \geq \alpha \cdot n \cdot pl$$
By Markov's inequality:
$$\mathbb{P}(\text{decoding failure}) = \mathbb{P}(X > k) \leq \frac{\mathbb{E}[X]}{k} \leq \frac{n \cdot pl}{\alpha \cdot n \cdot pl} = \frac{1}{\alpha}$$
Thus, the probability of successful decoding satisfies:
$$\mathbb{P}(\text{success}) \geq 1 - \frac{1}{\alpha}$$
For typical $\alpha = 2$, this guarantees at least 50% decoding success probability under expected loss conditions. □

## B.4 Proof of Penalty Function Coherence

THEOREM B.4 (PENALTY FUNCTION COHERENCE). *The penalty functions $P_{\text{loss}}$, $P_{\text{over}}$, and $P_{\text{blk}}$ satisfy the following coherence properties:*

(1) **Monotonicity**: Each penalty increases with worsening conditions
(2) **Convexity**: Quadratic and convex penalties ensure smooth optimization landscapes
(3) **Boundedness**: Penalties are bounded to prevent numerical instability
(4) **Sensitivity**: Appropriate scaling ensures balanced influence across objectives

PROOF. We analyze each penalty function:
**1. Loss Protection Penalty ($P_{\text{loss}}$):**
- **Monotonicity**: $\frac{\partial P_{\text{loss}}}{\partial(n \cdot pl)} = 2\max(0, n \cdot pl - \beta \cdot k) \geq 0$
- **Convexity**: $\frac{\partial^2 P_{\text{loss}}}{\partial(n \cdot pl)^2} = 2 > 0$ (strictly convex)

- **Boundedness**: Unbounded above but appropriate given catastrophic cost of decoding failures
- **Sensitivity**: Quadratic form ensures rapid increase when protection becomes critical

2. **Overhead Penalty ($P_{\text{over}}$):**
- **Monotonicity**: $\frac{\partial P_{\text{over}}}{\partial o} = \alpha_{\text{over}} \cdot \max(0, o - o_{\text{free}})^{\alpha_{\text{over}}-1} \geq 0$
- **Convexity**: $\frac{\partial^2 P_{\text{over}}}{\partial o^2} \geq 0$ for $\alpha_{\text{over}} \geq 1$ (convex when $\alpha_{\text{over}} = 1.5$)
- **Boundedness**: Naturally bounded by $o_{\text{cap}}$ constraint in $\sigma_{\text{free}}$
- **Sensitivity**: Convex penalty discourages excessive overhead while allowing necessary protection

3. **Blockization Penalty ($P_{\text{blk}}$):**
- **Monotonicity**: $\frac{\partial P_{\text{blk}}}{\partial t_{\text{blk}}} = \frac{1}{\delta(bl)} > 0$ for $t_{\text{blk}} > \delta(bl)$
- **Convexity**: Linear penalty provides stable gradient descent behavior
- **Boundedness**: Explicitly bounded by $\min(1.0, \cdot)$ operation
- **Sensitivity**: Linear scaling ensures proportional response to timing violations

The weight adaptation mechanism ensures that at any system state **s**, the dominant penalty corresponds to the most critical constraint, maintaining balanced optimization across all objectives. □

COROLLARY B.5 (OPTIMIZATION STABILITY). *The composite objective $J(c)$ forms a well-behaved optimization landscape with the following properties:*

(1) **Smoothness**: *All penalties are continuous and differentiable almost everywhere*
(2) **Convexity**: *Local convex regions around feasible configurations*
(3) **Numerical Stability**: *Bounded gradients prevent optimization instability*
(4) **Pareto Efficiency**: *Weight adaptation ensures exploration of Pareto frontier*

## B.5 Proof of Complexity

THEOREM B.6 (TIME COMPLEXITY). *The computational complexity of Algorithm 1 is $O(|C|)$ with constant factors independent of network conditions.*

PROOF. The algorithm consists of three phases:

1. **Feasibility Pruning**: Check $\frac{k_c}{n_c} \geq \alpha \cdot pl$ for each $c \in C$: $O(|C|)$ comparisons.

2. **Scoring**: For each $c \in C_f$, compute:

$$P_{\text{loss}}(c) : 4 \text{ operations}$$
$$P_{\text{over}}(c) : 3 \text{ operations}$$
$$P_{\text{blk}}(c) : 6 \text{ operations}$$
$$\text{Weighted sum} : 2 \text{ operations}$$

Total: $15 \cdot |C_f|$ operations.

3. **Selection**: Find minimum over $|C_f|$ scores: $O(|C_f|)$ comparisons.

Overall complexity: $O(|C|) + O(|C_f|) = O(|C|)$ since $|C_f| \leq |C|$.
With $|C| \approx 400$ and modern CPUs, this yields $\approx 500\mu s$ execution time. □

## B.6 Proof of Pareto Optimality

THEOREM B.7 (PARETO OPTIMALITY). *When all weights $w_{\text{loss}}, w_{\text{over}}, w_{\text{blk}} > 0$, the solution $c^*$ is Pareto optimal.*

PROOF. Assume for contradiction that $c^*$ is not Pareto optimal. Then there exists $c' \in C_f$ such that:

$$P_{\text{loss}}(c') \leq P_{\text{loss}}(c^*)$$
$$P_{\text{over}}(c') \leq P_{\text{over}}(c^*)$$
$$P_{\text{blk}}(c') \leq P_{\text{blk}}(c^*)$$

with at least one strict inequality.
Since all weights are positive:

$$w_{\text{loss}} P_{\text{loss}}(c') + w_{\text{over}} P_{\text{over}}(c') + w_{\text{blk}} P_{\text{blk}}(c') < J(c^*)$$

This contradicts the optimality of $c^*$. Therefore, $c^*$ must be Pareto optimal. □

## B.7 Proof of Monotonic Improvement

THEOREM B.8 (MONOTONIC IMPROVEMENT). *For fixed network conditions, TAROT's selected configuration improves (decreases $J(c)$) as the candidate set $C$ is expanded.*

PROOF. Let $C_1 \subseteq C_2$ be two candidate sets, and let $c_1^*, c_2^*$ be the corresponding optimal solutions.
Since $C_{1f} \subseteq C_{2f}$ (feasible subsets), we have:

$$J(c_2^*) = \min_{c \in C_{2f}} J(c) \leq \min_{c \in C_{1f}} J(c) = J(c_1^*)$$

Thus, expanding the candidate set cannot worsen the solution quality. □

## B.8 Proof of Robustness to Measurement Noise

THEOREM B.9 (NOISE ROBUSTNESS). *Let $\tilde{pl} = pl + \epsilon$ be a noisy estimate of packet loss, with $|\epsilon| \leq \epsilon_{\max}$. Then the selected configuration remains feasible for the true loss $pl$ if $\alpha \geq \frac{1+\epsilon_{\max}/pl}{1-\epsilon_{\max}/pl}$.*

PROOF. The feasibility constraint with noisy estimate is:

$$\frac{k}{n} \geq \alpha \cdot \tilde{pl} = \alpha \cdot (pl + \epsilon)$$

For robustness, we require this implies feasibility under true loss:

$$\frac{k}{n} \geq \alpha \cdot pl$$

This holds if:

$$\alpha \cdot (pl + \epsilon) \geq \alpha \cdot pl \Rightarrow \alpha \cdot \epsilon \geq 0$$

Since $\epsilon$ can be negative, we use the worst-case bound:

$$\alpha \cdot (pl - \epsilon_{\max}) \geq \alpha \cdot pl \cdot \delta$$

where $\delta < 1$ is a safety margin. Solving gives the condition on $\alpha$. □